\documentclass[12pt]{article}
\newcommand{\be}{\begin{equation}}
\newcommand{\ee}{\end{equation}}
\newcommand{\bea}{\begin{eqnarray}}
\newcommand{\eea}{\end{eqnarray}}
\newcommand{\nn}{\nonumber}

\newcommand{\bR}{{\bf R}}

\begin{document}
\baselineskip 3.4ex

\begin{center}
{\Huge \bf Theoretical and Numerical studies}\\
\bigskip
{\Huge \bf of the positions of cold trapped ions}\\
\bigskip\bigskip
{\small by }\\
\bigskip\bigskip
{\Large Todd P. Meyrath [a,b]}\\
\bigskip
{\small and }\\
\bigskip
{\Large Daniel F. V. James [c] \footnote {To whom correspondence should 
be addressed}}\\
\bigskip\bigskip
{\small [a] Box 332810, Georgia Institute of Technology,}\\
{\small  Atlanta, Ga 30332-1470, U.S.A.}\\
\bigskip
{\small [b] Los Alamos Summer School for AMO Physics}\\
{\small Los Alamos National Laboratory and 
University of New Mexico, Los Alamos}\\
{\small Los Alamos, NM 87544, U.S.A.}\\
\bigskip
{\small [c] Theoretical Division (T--4),  Mailstop B-268}\\
{\small University of California, 
Los Alamos National Laboratory, 
Los Alamos, NM 87545, U.S.A.}\\
{\small TEL: (505)-667-0956, FAX: (505)-665-3909, email: dfvj@t4.lanl.gov}\\

\bigskip
{submitted to \it Physics Letters A}\\
\bigskip
{\bf Abstract}\\
\end{center}
We examine the properties of cold ions confined by a 
Paul trap in a linear crystal configuration, a system
of considerable current interest due to its application
to practical quantum computation.  Using a
combination of theoretical and numerical calculation,
a semi-empirical formula for the positions of the ions
is derived.
\bigskip
\begin{center}
LA-UR-97-4771\\
{PACS numbers: 32.80.Pj, 42.50.Vk, 52.25.Wz, 03.67.Lx}\\
{Keywords: Ion Traps, Numerical simulations, experimental quantum 
computing}\\
\bigskip
\end{center}
\newpage
\noindent

Cold ions confined in electromagnetic traps
are rapidly becoming a very important
system both for the study of fundamental
physical systems, such as cold charged plasmas or
quantum chaos, and for technological applications
such as optical frequency standards.  Recently
a chain of cold ions in a linear trap has been proposed
as a possible means to realize a quantum computer \cite{CiracZoller}.
This idea was confirmed in principal soon after when an
elementary quantum logic gate was realized experimentally
using a trapped Beryllium ion \cite{NISTgate}, and larger
scale devices are currently being pursued by several experimental
groups (see, for example, \cite{LongPaper}).  
Understanding
the properties of collections of confined ions is of great
importance to these endeavors.

As is well known \cite{Earnshaw}, it is impossible to confine
charged particles by electrostatic forces alone.  To overcome
this problem, the radio-frequency Paul trap was developed: such
devices use a electromagnetic field varying at radio frequencies
($\sim$ 100 MHz say) to produce an {\em effective} binding potential
in three dimensions \cite{Ghosh}.  When two or more ions are confined
in such a trap, they will repel each other due to the Coulomb force.
As a result, such confined charged plasmas will have very low densities.
When sufficiently cold,
the plasma will condense into a crystalline state.  In the highly
anisotropic traps used for some atomic clocks \cite{hgclock} and for
quantum computing, this crystalline state 
is, for small enough numbers of ions, a simple chain of ions lying
in a straight line.  As the degree of anisotropy is decreased, or
number of ions is increased, phase changes to other configurations
will occur: firstly the ions adopt a zig-zag configuration, and
then a helical configuration.  These phase changes
have been studies numerically \cite{Schiffer} and analytically
\cite{Dubin}.

In this letter we present results of new numerical studies of
the positions of ions confined in highly anisotropic
traps.  This information is of course
of considerable importance in designing trapped ion quantum computers.
Using a simple theoretical argument we then develop
a relatively compact expression for the position
of each ion, which depends on the total number of ions confined in the
chain.  Our results are compared with both our numerical
data, and with results obtained previously by other authors,
and good agreement is obtained.

Consider a chain of $N$ ions confined in a linear trap (fig.1). The
position of the $n^{th}$ ion, where the ions are numbered from left
to right, will be denoted by the position vector relative to
the trap center (i.e. the minimum of the binding potential)
${\bf R}_{n}(t)=\left(X_n(t), Y_n(t), Z_n(t)\right)$.
The motion of each ion will be influenced by
an overall harmonic potential due to the trap electrodes,
and by the Coulomb force exerted
by all of the other ions.
Thus the potential energy of the ions in the ion chain is given by
the following expression
\bea
V\left(\bR_{1},\bR_{2},\ldots\bR_{N}\right)&=&
\frac{M}{2}\sum_{n=1}^{N}\left(
\omega_{x}^{2}X_{n}^{2}+
\omega_{y}^{2}Y_{n}^{2}+
\omega_{z}^{2}Z_{n}^{2}\right)\nn \\
&&+
\frac{e^2}{8\pi\epsilon_0}
\sum_{\stackrel{\scriptstyle n,m=1}{m \neq n}}^N
\frac{1}{|\bR_n-\bR_m|} \ ,
\label{potty}
\eea
where $M$ is the mass of each ion,  $e$ is the electron charge (the
ions are assumed to be singly ionized), 
$\epsilon_0$ is the permittivity of free space
and $\omega_x$ is the angular frequency characterizing the strength of the 
trapping potential in the x direction 
(and similarly $\omega_y$ and $\omega_z$
for the y and z directions). 

The equilibrium positions of the
ions, $\bR_{m}^{(0)}$
are defined by solutions of the following equations
\be
\left[\nabla 
V\left(\bR_{1},\bR_{2},\ldots\bR_{N}\right)\right]_{\bR_{m}
=\bR_{m}^{(0)}}=0
\label {equicond}
\ee
Substituting from (\ref{potty}) we obtain 
\be
M\omega_{i}^{2}X^{i(0)}_{m}-
\frac{e^2}{4\pi\epsilon_0}
\sum_{\stackrel{\scriptstyle n=1}{m \neq n}}^N
\frac{\left(X^{i(0)}_{n}-X^{i(0)}_{m}\right)}
{|\bR^{(0)}_n-\bR^{(0)}_m|^{3}}=0
\;\;(i=1,2,3) \ ,
\label{equicond2}
\ee
where $i=(1,2,3)$ denote the X, Y and Z components,
respectively.

Let us assume that the trap potentials are sufficiently 
strong in the Y and Z directions and sufficiently weak in
the X direction that in equilibrium the ions lie in a 
straight line along the X-axis.  Mathematically this
assumption is expressed by
\be
\bR^{(0)}_n=\ell \left(u_{n},0,0\right)
\label{druitt}
\ee
where $\ell$ is a scale length given by 
$\left(e^{2}/4\pi\epsilon_0 M 
\omega_{x}^{2}\right)^{1/3}$ and $u_{n}$ is
the dimensionless equilibrium position of the n-th
ion, which is a solution of
the following set of $N$ coupled algebraic equations,
obtained by substitution from eq(\ref{druitt}) into 
eq(\ref {equicond2}):
\be
u_m+
\sum_{\stackrel{\scriptstyle n=1}{m \neq n}}^N
\frac{{\rm sgn}(u_m-u_n)}{(u_m-u_n)^2}=0
\quad (m=1,2,\ldots N) \ ,
\label{ionequs}
\ee
where ${\rm sgn}(x)=1$ if $x>0$ and $-1$ if $x<0$.
For $N=2$ and $N=3$ these equations may be solved
analytically\cite{James}:
\bea
N=2:&\quad&u_1=-(1/2)^{2/3},
\quad u_2=(1/2)^{2/3} , \\
N=3:&\quad&u_1=-(5/4)^{1/3},
\quad u_2=0 ,
\quad u_3=(5/4)^{1/3} . 
\eea
For $N>2$ it is necessary to solve for the values
of $u_m$ numerically.
For small numbers of ions ($N\leq 100$ say) 
a Newton-Raphson method 
can be employed to find $u_m$; 
however this becomes inefficient
as N gets large. Therefore we used
another method, based on the following 
set of equations of motion
\be
\ddot{v}_{m}(\tau)=-\dot{v}_{m}(\tau)-v_m(\tau)+
\sum_{\stackrel{\scriptstyle n=1}{m \neq n}}^N
\frac{{\rm sgn}\left(v_m(\tau)-v_n(\tau)\right)}
{\left(v_m(\tau)-v_n(\tau)\right)^2}=0
\quad (m=1,2,\ldots N)  ,
\label{hypionequs}
\ee
where the single and double dots denote single and double 
differentiation with respect to the dimensionless time 
variable $\tau=\omega_{x}t$.  
These equations represent a hypothetical
damped oscillation of the ions in the trap, including their
mutual Coulomb interaction. The solutions of these
equations have the property that
\be
\lim_{\tau\rightarrow\infty}v_{m}(\tau)=u_{m} ,
\label{limpro}
\ee 
where $u_{m}$ are the desired solutions of  eq(\ref{ionequs}).
The integration of eq(\ref{hypionequs}) was 
carried out numerically using the standard 
fourth-order Runge-Kutta method\cite{Press}.
Because eq(\ref{limpro}) is valid regardless of
the initial conditions, the simplest possible initial
conditions were used, i.e. all of the ions being equally
spaced.  
The Runge-Kutta algorithm was applied repeatedly
until the values of $u_m$ between
adjacent iterations were identical to the seventh decimal place. 
This was done for up to 1000 ions (although not for all numbers).  
The values there by obtained are in agreement with those obtained
(for $N\leq 100$) by Newton-Raphson\cite{James}.  This dynamic
technique can be adapted quite easily to study classical wave
motion in the ion chain; this will be the subject of a forthcoming
paper.

In order to make some sense of the large amount of data generated
\footnote{ A data file called {\tt ion\_positions.dat} which 
contains the results of these numerical calculations can be 
found in the directory {\tt pub/james/Ion\_Position\_Data} 
which can be accessed via anonymous ftp to {\tt t4.lanl.gov}.},
we will now derive a analytic formula which approximates the numerical
results quite closely.  Our analysis is based on the very elegant idea
due to Garg \cite{garg}. Let us consider the force acting on an 
ion at position $X$.  The Coulomb force due to the two nearest 
neighbor ions is 
\bea
F_{nn}&=& \frac{e^{2}}{4\pi\epsilon_{0}}
\left(\frac{1}{S_{-}^{2}}-\frac{1}{S_{+}^{2}}\right) \nn \\
&\approx&\frac{e^{2}}{4\pi\epsilon_{0}}\frac{2}{S(X)^{2}}\frac{dS(X)}{dX}, 
\eea
where $S_{-}$ is the distance from the ion
to the nearest neighbor on the left,
$S_{+}$ is the distance to the nearest neighbor on the right and
$S(X)$ is the separation of ions at position $X$, treated as
a continuous function.  This is a reasonable approximation
to make for large numbers of ions.  The next 
nearest neighbors are approximately twice as far away as the nearest
neighbors, and so the force they exert on the ion is approximately
$F_{nn}/4$; the next pair of ions are three times as far away as the
nearest neighbors, and so the force they exert is approximately  
$F_{nn}/9$, and so on.  Thus the total Coulomb force on the ion will
approximately be given by the following expression:
\bea
F_{C}&\approx& F_{nn}\sum_{k=1}^{\infty} \frac{1}{k^{2}} \nn \\
&=&\frac{e^{2}}{4\pi\epsilon_{0}}\frac{\pi^{2}}
{3 S(X)^{2}}\frac{dS(X)}{dX}, 
\eea
where we have used the fact that $\sum_{k=1}^{\infty} 1/k^{2} = 
\pi^{2}/6$ and we have approximated the finite sum over all ions
as an infinite sum.  This approximation should be valid near the 
center of the ion chain, but will not yield very good results at
the ends of the chain.

The Coulomb force acting on the ion at position $X$ will be
balanced by the harmonic restoring force due to the trap electrodes.
Thus we can write the following identity:
\be
\frac{e^{2}}{4\pi\epsilon_{0}}\frac{\pi^{2}}{3 
S(X)^{2}}\frac{dS(X)}{dX}
-M\omega_{x}^{2}X=0 .
\label{crippen}
\ee
If we introduce the dimensionless ion separation $\sigma(u)=S(X)/\ell$
and  dimensionless distance from the trap center $u=X/\ell$,
where, as before 
$\ell=\left(e^{2}/4\pi\epsilon_0 M \omega_{x}^{2}\right)^{1/3}$, 
we obtain the following differential equation for the separation:
\be
\frac{d\sigma}{du}=\frac{3}{\pi^{2}}u\sigma^{2}.
\ee
This can be solved quite easily, yielding the formula \cite{garg}:
\be
\sigma(u)=\frac{2\pi^{2}/3}{C-u^{2}},
\label{cotton}
\ee
where $C$ is a constant, which could be determined from
the value of the separation of ions at the trap center ($u=0$).

Let $n(u)$ be the total number of ions which are within
a scaled distance $u$ of the trap center.  Clearly
$n(u)$ is given by the following differential equation
\be
\frac{d n(u)}{du}=\frac{1}{\sigma(u)}.
\ee
On substitution from eq(\ref{cotton}), and performing
the integration, we obtain the following formula
for $n(u)$:
\be
n(u)=A u -B u^{3},
\label{chapman}
\ee
where we have set $n(0)=0$.  The constants $A$ and $B$ can be related 
to the constant C introduced above.  However, instead of attempting 
to carry this analysis too far, it is better at this stage to obtain
empirical formulas for the the constants $A$ and $B$ based on our
numerical results.  This was done by performing a least squares fit of
the numerical data to a cubic formula of the type given by 
eq(\ref{chapman}).  The values of $A$ and $B$ were found for 
a variety of different total numbers of ions.  When this data
was compiled, we found that $A$ and $B$ were approximately 
given by the following
power laws:
\bea
A(N)&\approx& 0.436 N^{0.596} \nn \\
B(N)&\approx& 0.0375 N^{-0.178},  
\label{koslowski}
\eea
where $N$ is the total number of ions in the chain.

To obtain an expression for the position of the
n-th ion in the trap, it is necessary to invert eq(\ref{chapman}).
This can be done using the standard formulas for the roots
of a cubic equation \cite{spiegel}.  We therefore obtain,
taking care to select the correct root based on the value
of $n$ at $u=0$, the following formula for the the scaled
equilibrium positions of the n-th ion:
\bea
u_{n}&=&\sqrt{\frac{4A}{3B}}\cos\left(\frac{1}{3}
\cos^{-1}\left[
-\sqrt{\frac{27B}{4A^3}}\left\{n-\frac{(N+1)}{2}\right\} 
\right]+\frac{4\pi}{3}\right) \nn \\
&=&\alpha(N)\sin\left(\frac{1}{3}\sin^{-1}
\left[
\beta(N)
\left\{n-\frac{(N+1)}{2}\right\}\right]\right).
\label{deeming}
\eea
If we reintroduce the scale length $\ell$, we finally obtain
the following expression for the equilibrium position
of the $n-th$ ion, when there are a total of $N$ ions
in the trap:
\be
X^{(0)}_{n}=\left(\frac{e^{2}}{4\pi\epsilon_0 M 
\omega_{x}^{2}}\right)^{1/3}
\alpha(N)\sin\left(\frac{1}{3}\sin^{-1}
\left[
\beta(N)
\left\{n-\frac{(N+1)}{2}\right\}\right]\right),
\label{jacktheripper}
\ee
where, as before, the
ions are numbered from left to right,  
$M$ is the mass of each ion,  $e$ is the electron charge, 
$\epsilon_0$ is the permittivity of free space
and $\omega_x$ is the angular frequency characterizing 
the strength of the trapping potential in the x direction.
In eqs(\ref{deeming}) and (\ref{jacktheripper})
we have introduced the coefficients
$\alpha(N)=\sqrt{4A/3B}\approx 3.94 N^{0.387}$ and 
$\beta(N)=\sqrt{27B/4A^{3}}\approx 1.75 N^{-0.982}$.  

Equation (\ref{jacktheripper}) is the main result of this note.  
As an example
we have plotted in Figure 2 the numerically calculated
ion positions together with the positions calculated using
this formula, for a total of 41 ions in the trap.
Also we have included experimental ion position data
gleaned from Fig.5 of reference \cite{hgclock}. 
As can be seen from the figure, there is good agreement
between the numerical data and the empirical formula.
The differences between
the experimental data and that calculated 
numerically, may well be due to
the departure of the trapping potential from the harmonic
form we have assumed.  
The percentage r.m.s. error 
between the ion positions calculated
numerically and those calculated
using eq(\ref{deeming}) is shown in Figure 3.  The error is only
of the order of a few percent when $N>25$, but as expected,
errors increase for small numbers of ions.

For small arguments one can make the approximation
 $\sin(\sin^{-1}(x)/3)\approx x/3$, and
so, near the trap center [where $\left(n-{N+1}/2\right)$
is a small number] the scaled ion positions are given by:
\be
u_{n}\approx  2.29 \left(n-{N+1}/2\right) N^{-0.596} .
\ee
Hence the {\em minimum} separation between ions, which is
of considerable importance in quantum computer design \cite{HJKLP},
is given by 
\be
u_{min}(N)\approx  2.29 N^{-0.596} .
\label{warspite}
\ee
This result is in good agreement with the empirical
formulas previously calculated for the minimum separation of
ions are the trap center \cite{HJKLP}, \cite{Steane}. 
This formula is plotted in figure 4, along with the
numerical data, and the following formula for the
minimum ion spacing due
to Dubin \cite{Dubin} (see also \cite{garg}), based on
a fluid model for the ion cloud:
\be
u_{min}(N)\approx 1.92 N^{-2/3}\ln(aN)^{1/3},
\label{valiant}
\ee
where $a=6 e^{\gamma-13/5}\approx 0.794$, $\gamma$ being
Euler's constant.  As can been seen from figure 4, both
the empirical formula derived here
and the analytic formula due to Dubin
approximate the numerical data quite
closely.

\section*{Acknowledgments}
The authors would like to thank Anupam Garg and 
Albert Petschek for useful correspondence
and discussions.  This work was performed during the 
Los Alamos Summer School for Atomic, Molecular and 
Optical Physics, jointly funded by the 
Department of Energy Educational
programs and by the National Science Foundation
as an Research Experience for Undergraduates site
at the University of New Mexico; the authors
would like to thank Lee Collins, Norman Magee and 
Mike Zeilik for organizing it. This work was also
supported by the National Security Agency.

\newpage
\section*{Figure Captions}

\noindent
Figure 1. A schematic illustration of ions confined in an
harmonic trapping potential.

\vspace{10mm}
\noindent
Figure 2. Equilibrium positions of ions when there is a
total of 41 in the chain, as calculated numerically (crosses)
and by eq(\ref{deeming}) (plane line).  Also show are the
experimental positions of Hg+ ions in a linear trap, which
were gleaned from fig.5 of \cite{hgclock} (circles).

\vspace{10mm}
\noindent
Figure 3. Root mean square percentage error for calculating
positions of trapped ions using formula eq(\ref{deeming})
for total ion numbers $N$ up to 1000.

\vspace{10mm}
\noindent
Figure 4. Comparison of numerical results for minimum
ion separations (crosses) with the empirical power
law eq(\ref{warspite}) (plane line) and the analytic formula 
eq(\ref{valiant}) (dashed line).  The two curves are in
such good agreement that it is difficult to distinguish
them.


\begin{thebibliography}{99}

\bibitem{CiracZoller}  J. I. Cirac and  P. Zoller, Phys. Rev. Lett.  74, 
(1995) 4094-4097.

\bibitem{NISTgate}  C. Monroe, D. M. Meekhof, B. E. King,
W. M. Itano, and  D. J. Wineland,  Phys. Rev. Lett. 75 
(1995) 4714-4717.

\bibitem{LongPaper} R. J. Hughes, 
D. F. V. James, J. J. Gomez, M. S. Gulley,
M. H. Holzscheiter, P. G. Kwiat, S. K. Lamoreaux, C. G. Peterson,
V. D. Sandberg, M. M. Schauer, C. M. Simmons,
C. E. Thorburn, D. Tupa, P. Z. Wang, and  A. G. White,
``The Los Alamos Trapped Ion Quantum Computer Experiment'', 
Fortschritte der Physik, in the press, 1997

\bibitem{Earnshaw} S. Earnshaw, Trans. Cambridge Phil. Soc. 7 (1842), 97. 
See also, for instance, J. A. Stratton, Electromagnetic Theory 
(McGraw--Hill, New York, 1941), p. 116 ff.

\bibitem{Ghosh} P. K. Ghosh, Ion Traps (Clarendon Press, Oxford, 1995).

\bibitem{hgclock} M. E. Poitzsch, J. C. Bergquist, W. M. Itano
 and D. J. Wineland, Rev. Sci. Instrum. 67 (1996) 129-134.

\bibitem{Schiffer} J. P. Schiffer, Phys. Rev. Lett. 70 (1993) 818-821.

\bibitem{Dubin} D. H. E. Dubin, Phys. Rev. Lett. 71 (1993) 2753-2756.

\bibitem{James} D. F. V. James, ``Quantum dynamics of cold trapped ions,
with applications to quantum computation''
Applied Physics B: Lasers and Optics, in the press, 1997.

\bibitem{Press} W. H. Press, B. P. Flannery, 
S. A. Teukolsky, and W. T. Vetterling,
Numerical Recipies in C, $2^{nd}$ ed. (Cambridge University Press, 
New York, 1994).

\bibitem{garg} A. Garg, ``Vibrational Decoherence in Ion Trap Quantum 
Computers'', Proc. Conference on Fundamental Problems in Quantum 
Theory (held at University of Maryland Baltimore County, 4th-7th 
August, 1997), to appear; quant-ph/9710053. See Appendix B.

\bibitem{spiegel} M. R. Spiegel, Mathematical Handbook of Formulas 
and Tables, (McGraw-Hill, New York, 1968), eq(9.40), p.32.

\bibitem{HJKLP} R. J. Hughes, D. F. V. James, E. H. Knill,  
R. Laflamme and A. G. Petschek, Phys. Rev. Lett. 77 (1996) 3240-3243.

\bibitem{Steane}A. M. Steane, Applied Physics B 64 (1997) 623-642.


\end{thebibliography}
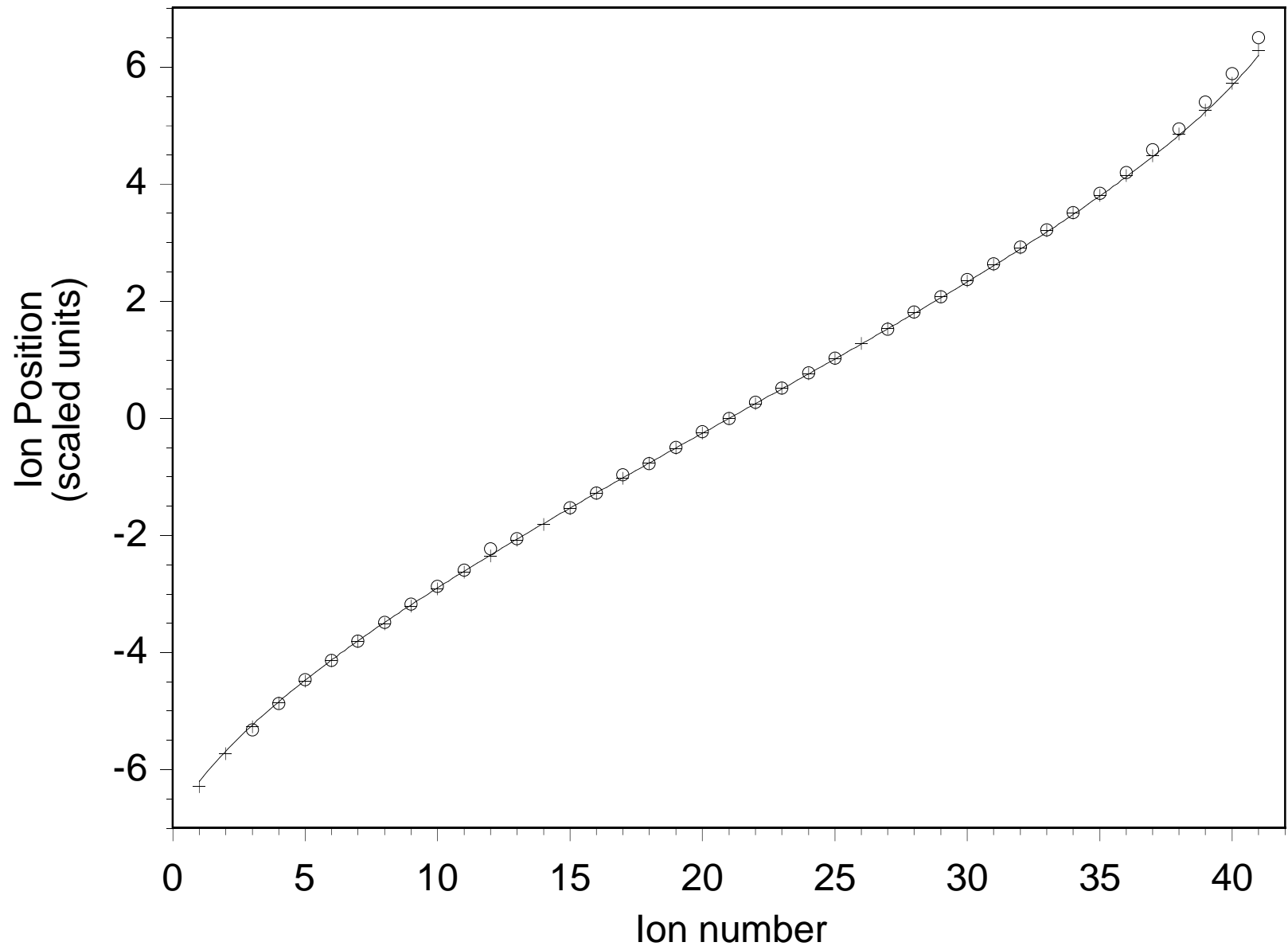
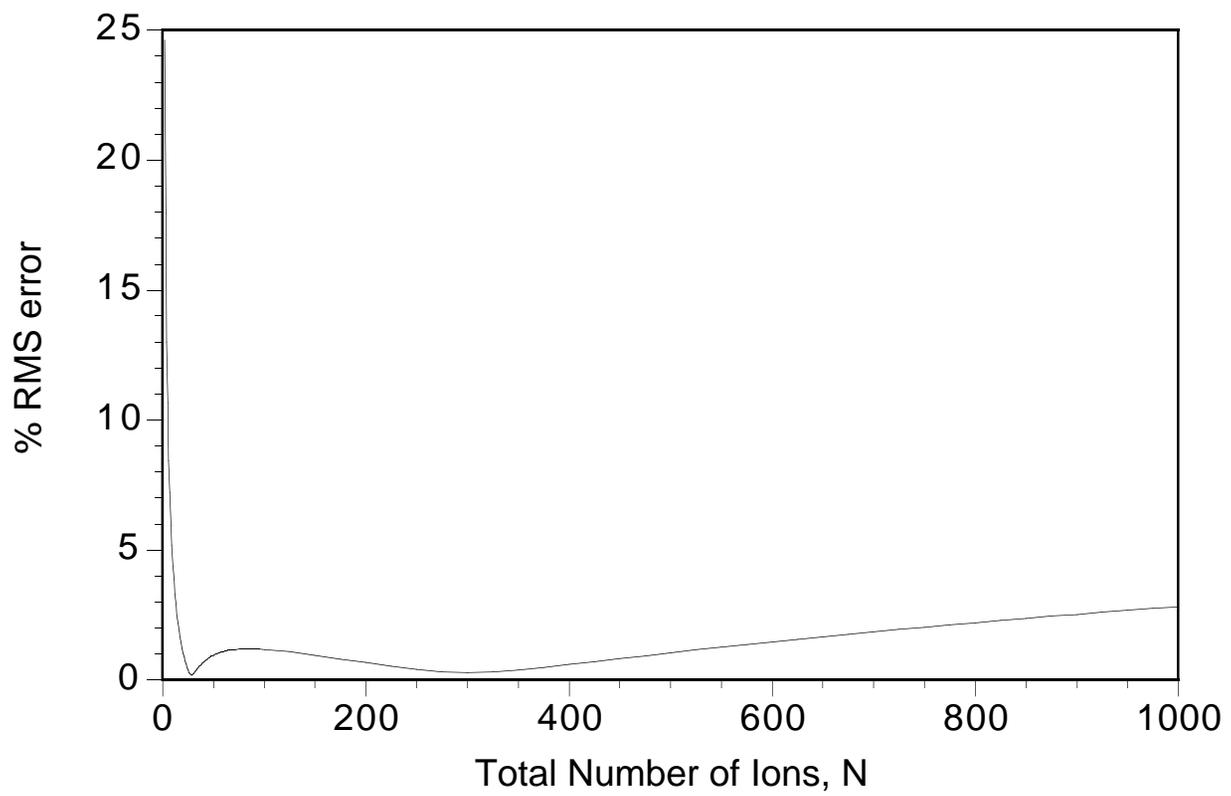
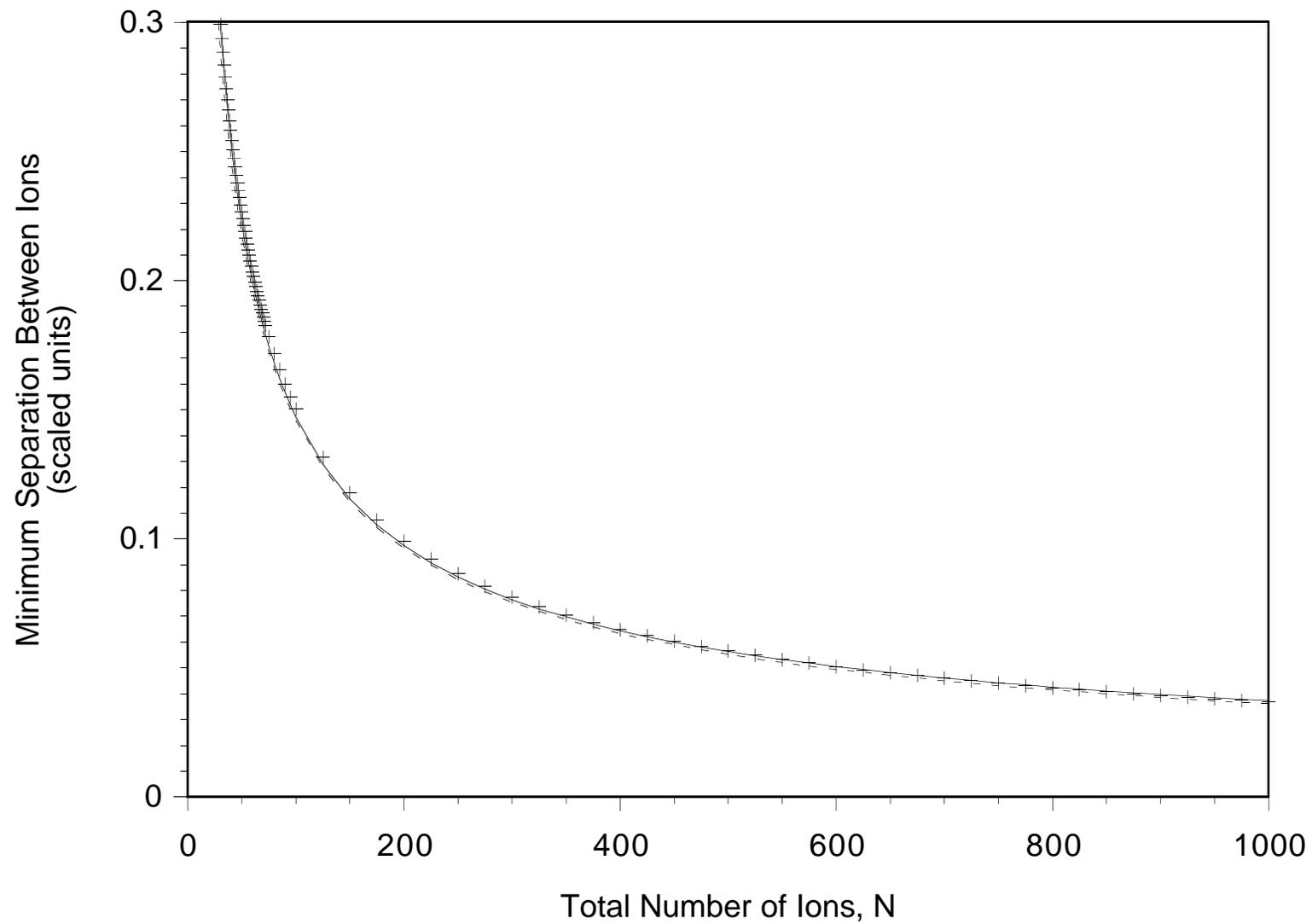
\end{document}